\documentclass[11pt]{article}

\usepackage[T1]{fontenc}
\usepackage{lmodern}
\usepackage{amsmath,amssymb,amsthm}
\usepackage[margin=1in]{geometry}
\usepackage{microtype}
\usepackage{booktabs}
\usepackage[hidelinks,pdfusetitle]{hyperref}
\usepackage{xcolor}

\newtheorem{theorem}{Theorem}[section]
\newtheorem{lemma}[theorem]{Lemma}
\newtheorem{proposition}[theorem]{Proposition}

\theoremstyle{definition}
\newtheorem{definition}[theorem]{Definition}
\theoremstyle{remark}

\newcommand{\Tr}{\operatorname{Tr}}
\newcommand{\id}{\operatorname{id}}
\newcommand{\Ad}{\operatorname{Ad}}
\newcommand{\supp}{\operatorname{supp}}
\newcommand{\ran}{\operatorname{ran}}

\newcommand{\ket}[1]{\lvert #1\rangle}
\newcommand{\bra}[1]{\langle #1\rvert}
\newcommand{\norm}[1]{\left\|#1\right\|_1}
\newcommand{\Z}{\mathcal Z_{\mathrm{cert}}}
\newcommand{\Ecert}{E_{\mathrm{cert}}}

\title{NPT states with zero distillable secret key}
\author{%
  Vjosa Blakaj\textsuperscript{1},
  Karol Horodecki\textsuperscript{2},
  Szymon Poszwa\textsuperscript{2},\\
  Thomas Theurer\textsuperscript{1},
  Marek Winczewski\textsuperscript{2}\\[0.7em]
  {\small\textsuperscript{1}Department of Mathematical Sciences,}\\
  {\small University of Copenhagen,}\\
  {\small Universitetsparken 5, 2100 Copenhagen \O, Denmark}\\[0.5em]
  {\small\textsuperscript{2}Institute of Informatics,
  National Quantum Information Centre,}\\
  {\small Faculty of Mathematics, Physics and Informatics,}\\
  {\small University of Gdansk,
  Wita Stwosza 57, 80-308 Gda\'nsk, Poland}
}

\hypersetup{
  pdfauthor={Vjosa Blakaj, Karol Horodecki, Szymon Poszwa,
             Thomas Theurer, Marek Winczewski}
}

\date{27 November 2026}

\begin{document}
\maketitle

\begin{abstract}
We prove that the Werner state $(23I-3F)/1448$ on
$\mathbb C^8\otimes\mathbb C^8$, with $F$ the SWAP operator,
has zero distillable secret key despite having negative partial
transpose. Our starting point is the rational matrix certificate
used by Faisal to establish that this state has zero distillable
entanglement. We extend the associated coherence bound to private
states with arbitrary finite shields. Every private-state test with
key dimension $d$ then accepts a normalized output of a separable
operation on any number of copies with probability at most
$(1+\sqrt d)/(2d)$.
This excludes secret-key distillation even when the shield grows
with the number of copies. We formulate the argument for a closed
convex family defined by auxiliary positivity conditions and show
that it is stable under tensor products of different members and
separable processing. Twice the regularized relative entropy to
this family bounds the distillable key of arbitrary input states;
for completed fixed-length protocols, rates above this threshold
obey an exponential strong converse. We verify the explicit
certificate in exact integer arithmetic and identify the precise
fixed-basis relaxation for which the private-test constant is optimal.
\end{abstract}

\section{Introduction}\label{sec:introduction}

Entanglement distillation asks whether Alice and Bob can turn many
copies of a shared state into maximally entangled pairs by local
operations and classical communication (LOCC). The discovery of
bound entanglement showed that an entangled mixed state need not
allow this conversion. In particular, an entangled state with
positive partial transpose (PPT) has zero distillable
entanglement~\cite{bound}. Whether negative partial transpose (NPT)
was sufficient for distillation became a longstanding question,
with evidence for NPT bound entanglement given in
Ref.~\cite{npt}.

Secret-key distillation asks a different question. Alice and Bob
seek a common random string that is independent of an eavesdropper,
Eve. The best asymptotic rate of secure bits per input copy is the
distillable key, denoted by $K_D$. Maximally entangled pairs yield
secret key, but states with zero distillable entanglement can also
have positive distillable key~\cite{secure,general}. Thus a proof
that a state cannot yield Bell pairs leaves open whether it can
still provide secrecy. This distinction is the starting point of
the present work.

The appropriate quantum targets for key distillation are private
states~\cite{general}. They consist of two key registers and
additional systems, called the shield, held by Alice and Bob.
Measuring the key registers gives matching, uniformly distributed
labels that are independent of Eve. Correlations with the shield
can protect this privacy even when entanglement distillation is
impossible. An obstruction to key distillation must therefore
control shields of arbitrary finite dimension, including shields
whose size increases with the number of input copies.

Faisal~\cite{faisal} supplied a rational matrix certificate for
the Werner state
\begin{equation}\label{eq:rho-star}
 \rho_*=\frac{23I-3F}{1448}
 \quad\text{on }\mathbb C^8\otimes\mathbb C^8.
\end{equation}
\textcolor{black}{Here $I$ denotes the identity operator on
$\mathbb C^8\otimes\mathbb C^8$, and $F$ is the SWAP operator,
defined by $F(\ket a\otimes\ket b)=\ket b\otimes\ket a$
for $\ket a,\ket b\in\mathbb C^8$.} This state is NPT:
if $\Gamma_B$ denotes transposition on Bob's system and
$\ket{\phi_8}=8^{-1/2}\sum_{i=0}^7\ket{ii}$, then
\begin{equation}\label{eq:intro-npt}
 \rho_*^{\Gamma_B}\ket{\phi_8}
 =-\frac1{1448}\ket{\phi_8}.
\end{equation}
Its negative partial transpose certifies that it is entangled.
The certificate in Ref.~\cite{faisal} nevertheless bounds the
Bell-state fidelity of every normalized two-qubit output obtained
from any number of copies under separable processing by a constant
strictly below one.
Hence, $\rho_*$ has zero distillable
entanglement.

We show that the same state has the stronger property
$K_D(\rho_*)=0$. Lemma~5 of Ref.~\cite{faisal} already provides
the required trace-norm inequality for four positive local
operators, including its extension to arbitrary copy number.
\textcolor{black}{We apply this inequality to bound the shield coherences relevant
to private-state distillation. These bounds rule out secret-key
distillation, which the Bell-state bound alone does not exclude.} This step excludes
secret key, which the Bell-state bound alone does not exclude.
The resulting example shows that negative partial transpose
does not guarantee distillable secrecy.

We also identify a family of states, denoted by $\Z$, for which
the argument applies. Here a certificate means an auxiliary
linear map, or equivalently its Choi matrix, satisfying explicit
positivity and matching  conditions. Section~\ref{sec:family}
states these conditions and includes their limits in trace norm.
The family contains the earlier local CP/coCP construction,
which gives PPT states with zero key~\cite{openai}, as well as the NPT state~\eqref{eq:rho-star}. We prove that members of
this larger family still have zero key when different members
are processed jointly. Relative entropy to the family then gives
a key bound for general input states, with an exponential strong
converse above twice its regularized value.

\subsection{Main result}

Let $K_A,K_B$ be key registers of dimension $d$, and let
$S=S_A S_B$ be a finite-dimensional shield. Write
\begin{equation}\label{eq:private-def}
 \ket{\phi_d}=\frac1{\sqrt d}\sum_{i=0}^{d-1}\ket{ii},
 \qquad \Phi_d=\ket{\phi_d}\bra{\phi_d}.
\end{equation}
A twisting is a unitary controlled by the key registers,
\begin{equation}\label{eq:twist}
 U=\sum_{a,b=0}^{d-1}\ket{ab}\bra{ab}\otimes U_{ab},
\end{equation}
where each $U_{ab}$ acts on the shield. For any shield state
$\tau_S$, it defines the private state
$\gamma_{d,U}=U(\Phi_d\otimes\tau_S)U^\dagger$~\cite{general}.
Measuring its key registers in the \textcolor{black}{computational} basis gives an ideal
secret key. The associated private-state test is~\cite{wilde}
\begin{equation}\label{eq:private-test}
 \Pi_{d,U}=U(\Phi_d\otimes I_S)U^\dagger.
\end{equation}
The pair $\{\Pi_{d,U},I-\Pi_{d,U}\}$ is a binary quantum
measurement whose first outcome accepts $\gamma_{d,U}$ with
probability one. It may require a joint measurement on Alice's
and Bob's systems. We use its acceptance probability to assess
the output of an LOCC protocol; the protocol itself need not
implement the test.

A separable completely positive (CP) map has product Kraus
operators, so it acts as
$\Lambda(X)=\sum_\ell(A_\ell\otimes B_\ell)
X(A_\ell\otimes B_\ell)^\dagger$.
Such maps include LOCC protocols and their selected outcomes.

\begin{theorem}[Uniform private-test bound]\label{thm:main}
Let $\rho\in\Z(A:B)$, with $\Z$ as in
Definition~\ref{def:certificate}. For every $n\ge1$, let
$\Lambda_n$ be \textcolor{black}{any} separable CP map with
$r_n=\Tr\Lambda_n(\rho^{\otimes n})>0$, and define its
normalized output by
$\omega_n=\Lambda_n(\rho^{\otimes n})/r_n$.
For every private-state test with key dimension $d\ge2$,
\begin{equation}\label{eq:main}
 \Tr(\Pi_{d,U}\omega_n)\le
 \beta_d:=\frac{1+\sqrt d}{2d}.
\end{equation}
The shield may have any finite dimension. Consequently,
$K_D(\rho)=0$ for every $\rho\in\Z$.
In particular, $\rho_*$ in~\eqref{eq:rho-star} belongs to
$\Z$, is NPT, and has zero distillable secret key.
\end{theorem}

\textcolor{black}{Here, } \emph{uniform} means that, for fixed $d$, the same
number $\beta_d$ applies for every copy number $n$, finite
shield, separable map, and twisting. In a distillation sequence,
the map, the shield, and the key dimension $d=d_n$ may all
depend on $n$.
For a trace-preserving protocol $r_n=1$. If $\Lambda_n$
describes a selected outcome, or collection of outcomes, of a
physical protocol, it is trace nonincreasing and $r_n$ is the
probability of that outcome. For a general CP map, $r_n$ is
simply the positive output trace used for normalization.

The theorem also applies to tensor products of different states
from $\Z$. Thus, joint processing of any finite collection of
members of this family cannot yield secret key. This conclusion
follows from the stability of their certificates under tensor
products, which we prove below.

\subsection{Intuition behind the result}

Correctness and secrecy impose two requirements on a distilled
key. Alice and Bob must obtain the same label, while Eve must
remain unable to distinguish the different possible labels.
In the private-state description, secrecy is reflected in
coherences between matching labels such as $\ket{ii}$ and
$\ket{jj}$. Each such coherence is an operator on the shield.
Its trace norm measures how much coherence remains when the
shield is taken into account.

For the states considered here, these coherences cannot be
separated from errors in the key labels. We show that a large
coherence between $\ket{ii}$ and $\ket{jj}$ requires nonzero
probability of the mismatching outcomes $\ket{ij}$ and
$\ket{ji}$. Suppressing the mismatches therefore also
suppresses the coherence needed for a private state. Summing
this constraint over all pairs of labels gives the bound
$\beta_d<1$. Since \textcolor{black}{the coherence bound} controls the trace norm of
the full shield operator, enlarging the shield cannot remove
the obstruction.

\subsection{Idea of the proof}

\textcolor{black}{The proof starts from the certificate conditions, which allow
us to bound the coherences needed for a private state.}
We first show that these conditions survive tensor
products and separable processing. \textcolor{black}{This allows the same argument to apply to any number of copies
and to joint processing of different states in $\Z$}

Next, let $\sigma$ be a state in this family. Choose positive
operators $X_0,X_1$ on Alice's system and $Y_0,Y_1$ on Bob's,
and define $D_{ij}=X_i\otimes Y_j$ and
$p_{ij}=\Tr(D_{ij}\sigma)$ for $i,j\in\{0,1\}$.
The resulting coherence bound is
\begin{equation}\label{eq:roadmap-four}
 \norm{\sqrt{D_{00}}\,\sigma\sqrt{D_{11}}}
 \le(p_{00}p_{01}p_{10}p_{11})^{1/4}.
\end{equation}
Thus, the coherence between the two matching choices is bounded
by all four weights, including the crossed choices. We give
the four-effect argument of Ref.~\cite{faisal} for the whole
family, starting from the weaker estimate of Ref.~\cite{openai}.
The initial estimate contains constant prefactors. Applying it
to tensor powers and then taking roots removes those prefactors
and gives~\eqref{eq:roadmap-four}.

Finally, consider the output of a separable operation. The matrix
block between $\ket{ii}$ and $\ket{jj}$ is an operator on
the shield. We relate its
trace norm to the left-hand side of~\eqref{eq:roadmap-four}
and show that the bound survives summation over the operation's
Kraus terms. Summing over the key labels then bounds the
private-test acceptance. A private state is accepted with
probability one, so the gap below one prevents the protocol's
output from approaching any private state.

Sections~\ref{sec:family}--\ref{sec:key} give this argument.
Section~\ref{sec:explicit} inserts the explicit certificate
of Ref.~\cite{faisal}, and Section~\ref{sec:entropy} derives
the relative-entropy bound and strong converse. The appendices
provide the exact certificate verification, the passage from
classical key to private-state tests, and the sharpness result
for the fixed-basis shield constraints.

\section{A family defined by auxiliary certificates}\label{sec:family}

\subsection{Conventions and definition}

All Hilbert spaces are finite dimensional, and $\mathcal B(H)$
denotes the linear operators on $H$. Logarithms are to base two,
and $\norm{X}=\Tr\sqrt{X^\dagger X}$ is the trace norm.
We use Greek letters for normalized states and $R$ for a positive
operator whose trace need not be one.

\textcolor{black}{Fix a computational basis on each local space and use the
corresponding product bases on composite spaces.} The map $T_H$ is the full
transpose on $\mathcal B(H)$, so $T_H(X)=X^T$; if
$H=H_A\otimes H_B$, it transposes both factors.
\textcolor{black}{A linear map $\mathcal N:\mathcal B(H)\to\mathcal B(K)$
is completely copositive, or coCP, if $\mathcal N\circ T_H$
is completely positive.}
For $\mathcal C:\mathcal B(H)\to\mathcal B(K)$, our Choi convention is
\begin{equation}\label{eq:choi}
 J_{\mathcal C}=\sum_{a,b}\ket a\bra b\otimes
                    \mathcal C(\ket a\bra b),
\end{equation}
\textcolor{black}{where $J_{\mathcal C}\in\mathcal B(H\otimes K)$.} For a Choi matrix, $\Gamma$ denotes
partial transpose on the entire input factor $H$, leaving $K$
unchanged. For a bipartite state, $\Gamma_B$ denotes partial
transpose on Bob's share. We retain subsystem indices when
several different partial transposes occur in the same calculation.

A certified representation describes a state by two local CP maps
acting on an auxiliary pure input. We call this input the
\emph{ancestor}. 
\textcolor{black}{It is part of the representation and is not supplied as a
resource to the distillation protocol.}
The representing maps may change the trace, and the ancestor
need not be normalized. Any nonzero positive output $R$ defines
the normalized state $R/\Tr R$.

\color{black}
We use $A$ and $B$ to denote the Hilbert spaces of Alice's
and Bob's output systems. Thus $\mathcal B(A)$ is the space
of linear operators on Alice's output space, and similarly
for $\mathcal B(B)$. The spaces $H_A$ and $H_B$ are auxiliary
input spaces used in the representation. Their dimensions
need not coincide with those of $A$ and $B$, respectively.
\color{black}

\begin{definition}[Certified representation]\label{def:certificate}
Let $H=H_A\otimes H_B$, let $\ket\chi\in H$, and let
$L:\mathcal B(H_A)\to\mathcal B(A)$ and
$M:\mathcal B(H_B)\to\mathcal B(B)$ be CP maps.
Put $\mathcal C=L\otimes M$. The representation
\begin{equation}\label{eq:representation}
 R=\mathcal C(\ket\chi\bra\chi)
\end{equation}
is certified if there \textcolor{black}{exists} a Hermiticity-preserving map $\mathcal S$
such that
\begin{equation}\label{eq:map-certificate}
 \mathcal C+\mathcal S\ \text{and}\ \mathcal C-\mathcal S
 \ \text{are CP},\qquad
 \mathcal S\circ T_H\ \text{is CP},
\end{equation}
and
\begin{equation}\label{eq:state-match}
 \mathcal S(\ket\chi\bra\chi)=\mathcal C(\ket\chi\bra\chi).
\end{equation}
Let $\mathcal G^+_{A:B}$ consist of all such positive outputs,
allowing arbitrary finite ancestor dimensions and unnormalized
ancestors. Let $\mathcal G_{A:B}$ consist of those outputs with
trace one. Finally, $\Z(A:B)$ is the closure of
$\mathcal G_{A:B}$ in trace norm. Explicitly,
\begin{equation}\label{eq:family}
 \rho\in\Z(A:B)
 \quad\Longleftrightarrow\quad
 \exists\,(\rho_k)_{k\ge1}\subset\mathcal G_{A:B}
 \text{ such that }\lim_{k\to\infty}\norm{\rho_k-\rho}=0.
\end{equation}
\end{definition}

The local CP maps $L$ and $M$ generate the positive output.
The auxiliary map $\mathcal S$ supplies the certificate through
\eqref{eq:map-certificate} and~\eqref{eq:state-match}; it need
not itself be CP. These conditions can be checked as matrix
positivity and kernel conditions, as shown next. They do not
require a rational certificate in general. Rational entries are
used for the explicit example in Section~\ref{sec:explicit}.
\textcolor{black}{We do not assume that every state in $\Z$ admits a certified
representation with finite-dimensional auxiliary input spaces.}

\color{black}
Since $\mathcal C-\mathcal S$ is CP, equality of the two maps
on $\ket\chi\bra\chi$ implies equality on $\ket z\bra\chi$
and $\ket\chi\bra z$ for every $z\in H$.
\color{black}

\begin{lemma}[Equivalent matching conditions]\label{lem:matching}
Under~\eqref{eq:map-certificate}, condition~\eqref{eq:state-match}
is equivalent to
\begin{equation}\label{eq:row-match}
 (\mathcal C-\mathcal S)(\ket z\bra\chi)=
 (\mathcal C-\mathcal S)(\ket\chi\bra z)=0
 \quad\text{for every }z\in H.
\end{equation}
\textcolor{black}{Write $J=J_{\mathcal C}$.
The representation~\eqref{eq:representation} is certified
if and only if there exists a Hermitian operator
$Q\in\mathcal B(H\otimes A\otimes B)$ such that}
\begin{align}
 J+Q&\succeq0,\qquad J-Q\succeq0,\qquad
 Q^{\Gamma}\succeq0,\label{eq:choi-positive}\\
 &(J-Q)(\ket{\bar\chi}\otimes I_{AB})=0.\label{eq:choi-match}
\end{align}
\textcolor{black}{The corresponding auxiliary map $\mathcal S$ is uniquely determined
by $J_{\mathcal S}=Q$.} Here the bar denotes componentwise conjugation in the chosen input
basis. \textcolor{black}{For Hermitian operators $X$ and $Y$, we write $X\succeq Y$
if $X-Y$ is positive semidefinite.}
\end{lemma}

\begin{proof}
Write the CP map $\mathcal D=\mathcal C-\mathcal S$ in Kraus form,
$\mathcal D(X)=\sum_\mu K_\mu X K_\mu^\dagger$.
Equation~\eqref{eq:state-match} gives
$\sum_\mu\ket{K_\mu\chi}\bra{K_\mu\chi}=0$, so
$K_\mu\chi=0$ for every $\mu$. This proves~\eqref{eq:row-match};
the converse follows by taking $z=\chi$.
\textcolor{black}{The Choi correspondence is a bijection between
Hermiticity-preserving linear maps and Hermitian operators.}
The three positivity conditions in~\eqref{eq:choi-positive}
are the Choi forms of~\eqref{eq:map-certificate}.
\color{black}
Finally, write $\ket\chi=\sum_b\chi_b\ket b$ in the chosen
input basis. The operator $\ket{\bar\chi}\otimes I_{AB}$
maps each $y\in A\otimes B$ to $\ket{\bar\chi}\otimes y$.
Since $J-Q=J_{\mathcal D}$, expanding~\eqref{eq:choi} gives
\begin{align*}
 (J-Q)(\ket{\bar\chi}\otimes y)
 &=\sum_{a,b}\overline{\chi_b}\,
   \ket a\otimes\mathcal D(\ket a\bra b)y\\
 &=\sum_a\ket a\otimes\mathcal D(\ket a\bra\chi)y.
\end{align*}
The second equality uses
$\bra\chi=\sum_b\overline{\chi_b}\bra b$
and the linearity of $\mathcal D$.

Because the input basis is orthonormal, applying
$\bra a\otimes I_{AB}$ to the last expression gives
$\mathcal D(\ket a\bra\chi)y$.
Thus~\eqref{eq:choi-match} holds if and only if
\[
 \mathcal D(\ket a\bra\chi)y=0
 \qquad\text{for every }a\text{ and every }y\in A\otimes B.
\]
Since $y$ is arbitrary, this is equivalent to
$\mathcal D(\ket a\bra\chi)=0$ for every input basis vector
$\ket a$. By linearity, it is also equivalent to
$\mathcal D(\ket z\bra\chi)=0$ for every $z\in H$.
Moreover, $\mathcal D$ preserves Hermiticity, so
\[
 \mathcal D(\ket\chi\bra z)
 =\mathcal D(\ket z\bra\chi)^\dagger.
\]
Hence the adjoint condition holds as well.
We have therefore shown that~\eqref{eq:choi-match}
is equivalent to~\eqref{eq:row-match}, which the first part
of the proof showed to be equivalent to~\eqref{eq:state-match}.
\color{black}
\end{proof}

The two uses of partial transpose concern different systems.
The condition $Q^\Gamma\succeq0$ transposes the full input
of the auxiliary map, whereas the PPT condition on the state
transposes only Bob's output. The certificate therefore does
not require the represented state to be PPT. This distinction
allows the family to contain the NPT example.

\subsection{Stability under mixing and processing}

\textcolor{black}{We next show that certified outputs remain certified under
mixing, tensor products, and separable CP maps.} Both signs in~\eqref{eq:map-certificate}
matter: the difference gives the restricted block positivity, whereas
the sum allows that difference to remain CP under tensor products.

\begin{proposition}[Closure properties]\label{prop:closure}
The set $\mathcal G^+$ is a convex cone, is closed under tensor
products of different members, and is preserved by separable CP maps.
Consequently, $\Z$ is a closed convex family, is closed under
mixed tensor products, and is preserved by normalized nonzero
separable outputs. It contains every separable state.
\end{proposition}

\begin{proof}
We first prove the assertions for certified representations, and then
pass to their trace-norm closure.

For tensor products, we group the input and output tensor factors
by party. Take two certified triples
$(\mathcal C_i,\mathcal S_i,\chi_i)$, where
$\mathcal C_i=L_i\otimes M_i$ acts on
$H_i=H_{A,i}\otimes H_{B,i}$, and write
$R_i=\mathcal C_i(\ket{\chi_i}\bra{\chi_i})$. Set
\[
 \mathcal C_{12}=\mathcal C_1\otimes\mathcal C_2,
 \qquad
 \mathcal S_{12}=\mathcal S_1\otimes\mathcal S_2,
 \qquad
 \ket{\chi_{12}}=\ket{\chi_1}\otimes\ket{\chi_2}.
\]
The maps
$\mathcal A_i=(\mathcal C_i+\mathcal S_i)/2$ and
$\mathcal B_i=(\mathcal C_i-\mathcal S_i)/2$ are CP. Hence
\begin{align}
 \mathcal C_{12}+\mathcal S_{12}
 &=2(\mathcal A_1\otimes\mathcal A_2
      +\mathcal B_1\otimes\mathcal B_2),\notag\\
 \mathcal C_{12}-\mathcal S_{12}
 &=2(\mathcal A_1\otimes\mathcal B_2
      +\mathcal B_1\otimes\mathcal A_2)
 \label{eq:tensor-domination}
\end{align}
are CP. Moreover,
\[
 \mathcal S_{12}\circ T_{H_1\otimes H_2}
 =(\mathcal S_1\circ T_{H_1})
   \otimes(\mathcal S_2\circ T_{H_2})
\]
is CP, and agreement with $\mathcal C_{12}$ holds on
$\ket{\chi_{12}}\bra{\chi_{12}}$ by agreement in each factor.
The local maps are $L_1\otimes L_2$ and $M_1\otimes M_2$.
We therefore obtain a certified representation of $R_1\otimes R_2$,
even when the two factors are different.

We next consider a finite nonnegative combination $\sum_k t_kR_k$
on a fixed output space $A:B$. Choose certified representations
$R_k=(L_k\otimes M_k)(\ket{\chi_k}\bra{\chi_k})$ with
auxiliary maps $\mathcal S_k$, and write
$\mathcal C_k=L_k\otimes M_k$ and
$H_k=H_{A,k}\otimes H_{B,k}$. Introduce
\[
 H_A=\bigoplus_k H_{A,k},\qquad
 H_B=\bigoplus_k H_{B,k},\qquad H=H_A\otimes H_B,
\]
with canonical embeddings $V_k:H_{A,k}\to H_A$ and
$W_k:H_{B,k}\to H_B$. Define
\begin{align}
 \ket\chi
 &=\sum_k\sqrt{t_k}(V_k\otimes W_k)\ket{\chi_k},\notag\\
 L(X)&=\sum_k L_k(V_k^\dagger XV_k),\qquad
 M(Y)=\sum_k M_k(W_k^\dagger YW_k).
 \label{eq:direct-sum}
\end{align}
Both local maps are CP. To check the certificate on the whole
input space, including the sectors with different indices, let
\[
 \mathcal P_{kl}(Z)
 =(V_k\otimes W_l)^\dagger Z(V_k\otimes W_l).
\]
Each $\mathcal P_{kl}$ is a CP compression map. For
$\mathcal C=L\otimes M$, define the auxiliary map by
\[
 \mathcal S=\sum_k\mathcal S_k\circ\mathcal P_{kk}.
\]
Expanding the product $L\otimes M$ gives
\[
 \mathcal C\pm\mathcal S
 =\sum_k(\mathcal C_k\pm\mathcal S_k)\circ\mathcal P_{kk}
  +\sum_{k\ne l}(L_k\otimes M_l)\circ\mathcal P_{kl}.
\]
Every term on the right is CP, so $\mathcal C+\mathcal S$
and $\mathcal C-\mathcal S$ are CP on the whole input space.
Since the canonical embeddings are real in the direct-sum bases,
\[
 \mathcal P_{kk}\circ T_H=T_{H_k}\circ\mathcal P_{kk},
 \qquad
 \mathcal S\circ T_H
 =\sum_k(\mathcal S_k\circ T_{H_k})\circ\mathcal P_{kk}
\]
is CP as well. The compression of $\ket\chi\bra\chi$ to a matched
sector is $t_k\ket{\chi_k}\bra{\chi_k}$, and its compression to
an unmatched sector is zero. Therefore
\[
 \mathcal C(\ket\chi\bra\chi)
 =\mathcal S(\ket\chi\bra\chi)
 =\sum_k t_kR_k.
\]
\textcolor{black}{Thus any finite nonnegative linear combination of certified
outputs is again certified}. This proves that $\mathcal G^+_{A:B}$ is a convex cone. All auxiliary
spaces used here are finite dimensional, as allowed by the definition.

Now let $\Lambda$ be a separable CP map from $A:B$ to $A':B'$.
In finite dimensions it has a finite product-Kraus representation
\[
 \Lambda(X)=\sum_\ell K_\ell X K_\ell^\dagger,
 \qquad K_\ell=A_\ell\otimes B_\ell.
\]
For a certified output $R=\mathcal C(\ket\chi\bra\chi)$ with
$\mathcal C=L\otimes M$ on an ancestor space $H$,
use $\Ad_V(X)=VXV^\dagger$ to define
\[
 L_\ell'=\Ad_{A_\ell}\circ L,\qquad
 M_\ell'=\Ad_{B_\ell}\circ M,\qquad
 \mathcal S_\ell'=\Ad_{K_\ell}\circ\mathcal S.
\]
Then $\mathcal C_\ell'=L_\ell'\otimes M_\ell'
=\Ad_{K_\ell}\circ\mathcal C$, and
\[
 \mathcal C_\ell'\pm\mathcal S_\ell'
 =\Ad_{K_\ell}\circ(\mathcal C\pm\mathcal S),
 \qquad
 \mathcal S_\ell'\circ T_H
 =\Ad_{K_\ell}\circ(\mathcal S\circ T_H)
\]
are CP. Agreement on $\ket\chi\bra\chi$ is also preserved.
Thus $K_\ell R K_\ell^\dagger\in\mathcal G^+_{A':B'}$ for
each term, and the cone property gives
$\Lambda(R)\in\mathcal G^+_{A':B'}$. The direct-sum construction
above supplies a certified product representation of this sum.

\color{black}
Let $P\succeq0$ on $A$ and $Q\succeq0$ on $B$.
Choose one-dimensional auxiliary input spaces $H_A,H_B$,
put $H=H_A\otimes H_B$, and take a unit vector
$\ket\chi\in H$. Define
\[
 L(X)=\Tr(X)P,\qquad M(Y)=\Tr(Y)Q,
 \qquad \mathcal C=L\otimes M.
\]
Both local maps are CP, and
$\mathcal C(\ket\chi\bra\chi)=P\otimes Q$.
Choose $\mathcal S=\mathcal C$. Since $H$ is one-dimensional,
the transpose $T_H$ is the identity map, so
\[
 \mathcal C+\mathcal S=2\mathcal C,\qquad
 \mathcal C-\mathcal S=0,\qquad
 \mathcal S\circ T_H=\mathcal C.
\]
All three maps are CP, including the zero map.
The matching condition holds because $\mathcal S=\mathcal C$.
Thus $P\otimes Q$ admits a certified representation.
Taking finite sums and using the cone property proved above
shows that $\mathcal G^+$ contains every separable positive
operator. Its trace-one slice $\mathcal G$ therefore contains
every separable state.
\color{black}

Finally, $\mathcal G$ is convex because it is the trace-one slice
of a convex cone. Its trace-norm closure $\Z$ is therefore closed
and convex. To check tensor stability of the closure, choose
$\rho_j,\sigma_j$ in the respective trace-one represented families
such that $\rho_j\to\rho$ and $\sigma_j\to\sigma$ in trace norm.
The product $\rho_j\otimes\sigma_j$ is represented, and
\[
 \norm{\rho_j\otimes\sigma_j-\rho\otimes\sigma}
 \le\norm{\rho_j-\rho}+\norm{\sigma_j-\sigma}
 \longrightarrow0.
\]
Hence $\rho\otimes\sigma$ belongs to the corresponding closed family.

For normalized processing, fix a separable CP map $\Lambda$ and
$\rho\in\Z(A:B)$ with $r=\Tr\Lambda(\rho)>0$.
Choose $\rho_j\in\mathcal G_{A:B}$ converging to $\rho$, and set
$R_j=\Lambda(\rho_j)$, $R=\Lambda(\rho)$, and $r_j=\Tr R_j$.
Continuity gives $R_j\to R$ and $r_j\to r$, so $r_j>0$ for all
sufficiently large $j$. The cone property then gives
$R_j/r_j\in\mathcal G_{A':B'}$, while positivity implies
\[
 \norm{\frac{R_j}{r_j}-\frac Rr}
 \le\frac{\norm{R_j-R}+|r_j-r|}{r}
 \le\frac{2}{r}\norm{R_j-R}
 \longrightarrow0.
\]
It follows that $\Lambda(\rho)/\Tr\Lambda(\rho)\in\Z(A':B')$.
These limit arguments concern states on fixed output spaces and require
no common bound on the ancestor dimensions. Together with the inclusion
of separable states proved above, they establish all the assertions.
\end{proof}

The local CP/coCP family is a subfamily of $\mathcal G^+$.
Indeed, if both $L$ and $M$ are CP and coCP, one may choose
$\mathcal S=\mathcal C$. These outputs are PPT, since
$T_B\circ M$ is CP. Their trace-norm closure is therefore PPT
as well. The example in Section~\ref{sec:explicit} will show
that the auxiliary certificate gives a strict extension.

\section{From the certificate to a coherence bound}\label{sec:effects}

\subsection{Restricted block positivity}

We use the restricted-positivity argument of Ref.~\cite{faisal}
for the general representations of Definition~\ref{def:certificate}.
Complete positivity of $\mathcal C$ gives positive block matrices
with the usual ordering of the off-diagonal terms. The certificate
provides another ordering when the two vectors add up to the
ancestor. That restricted statement is exactly what the proof needs.

\begin{lemma}[Restricted block positivity]\label{lem:block}
Let $(\mathcal C,\mathcal S,\chi)$ be certified.
For any $v,w\in H$ with $v+w=\chi$,
\begin{equation}\label{eq:block}
 \begin{pmatrix}
  \mathcal C(\ket w\bra w)&\mathcal C(\ket v\bra w)\\
  \mathcal C(\ket w\bra v)&\mathcal C(\ket v\bra v)
 \end{pmatrix}\succeq0.
\end{equation}
\end{lemma}

\begin{proof}
The rank-one block matrix formed from the vector $(\chi,z)$ is
positive. Transpose the entire matrix, including its two-dimensional
flag, and apply the CP map
$\id_2\otimes(\mathcal S\circ T_H)$. The result is
\begin{equation}\label{eq:auxiliary-block}
 \begin{pmatrix}
  \mathcal S(\ket\chi\bra\chi)&\mathcal S(\ket z\bra\chi)\\
  \mathcal S(\ket\chi\bra z)&\mathcal S(\ket z\bra z)
 \end{pmatrix}\succeq0.
\end{equation}
Lemma~\ref{lem:matching} replaces the first three entries by
those of $\mathcal C$. Adding the positive operator
$(\mathcal C-\mathcal S)(\ket z\bra z)$ to the last diagonal
entry replaces that entry as well. Set $z=v$ and apply the
congruence $B\mapsto TBT^\dagger$ with
$T=\left(\begin{smallmatrix}I&-I\\0&I\end{smallmatrix}\right)$.
Since $\chi-v=w$, the resulting matrix is~\eqref{eq:block}.
\end{proof}

Proposition~\ref{prop:closure} supplies the same statement for every
mixed tensor product of certified representations. No assumption
about orthogonality of $v$ and $w$ enters the argument.

\subsection{Separating the two local effects}

\begin{lemma}[Initial four-effect estimate]\label{lem:weak-four}
Let $(\mathcal C,\mathcal S,\chi)$ be a certified representation
and put $R=\mathcal C(\ket\chi\bra\chi)$.
For positive local operators $X_0,X_1,Y_0,Y_1$, define
\begin{equation}\label{eq:four-data}
 D_{ij}=X_i\otimes Y_j,\qquad
 p_{ij}=\Tr(D_{ij}R),\qquad
 x=\norm{\sqrt{D_{00}}R\sqrt{D_{11}}}.
\end{equation}
Then
\begin{equation}\label{eq:weak-four}
 x\le\sqrt{p_{00}p_{01}}+\sqrt{p_{10}p_{11}}
       +2\sqrt{p_{01}p_{10}}.
\end{equation}
\end{lemma}

\begin{proof}
Pull the effects back to the ancestor: define
$E_i=L^\dagger(X_i)$ and $F_j=M^\dagger(Y_j)$.
Then, $E_i, F_j$ are positive semi-definite. Let $H=E_0+E_1$, and define the positive
invertible operator $N$ to equal $H^{-1/2}$ on $\supp H$
and the identity on $\ker H$. The operators
$E_i'=NE_iN$ commute, because their sum is the
orthogonal projection onto $\supp H$ and each is supported there.
Let $P$ be the nonnegative spectral projection of $E_0'-E_1'$.

We use this spectral split to decompose the ancestor:
\begin{equation}\label{eq:oblique-split}
 v=(NPN^{-1}\otimes I)\chi,\qquad
 w=(N(I-P)N^{-1}\otimes I)\chi.
\end{equation}
The operator $NPN^{-1}$ is an idempotent, generally not an
orthogonal projection. Its role is to give $v+w=\chi$ while
keeping the original certified maps unchanged.
Write $R_{ab}=\mathcal C(\ket a\bra b)$ for
$a,b\in\{v,w\}$, and put
$a_{ij}=\Tr(D_{ij}R_{vv}) = \bra{v}E_i \otimes F_j \ket{v}$ and
$b_{ij}=\Tr(D_{ij}R_{ww})=\bra{w}E_i \otimes F_j \ket{w} $.
Since $PE'_i = E'_i P$:
\begin{equation}\label{eq:crossed-trace}
 \Tr(D_{ij}R_{vw})
 =\bra\chi\bigl[N^{-1}(I-P)E_i'PN^{-1}\otimes F_j\bigr]\ket\chi
 =0.
\end{equation}
Similarly, $\Tr(D_{ij}R_{wv})=0$.
The spectral definition of $P$, and expanding $\ket{\chi}\bra{\chi}$, where $\chi = v + w$ gives that
\begin{equation}\label{eq:split-weights}
 p_{ij}=a_{ij}+b_{ij},\qquad
 a_{0j}\ge a_{1j},\qquad b_{0j}\le b_{1j}.
\end{equation}
The two inequalities above follow by expanding the definitions of \(a_{ij}\) and \(b_{ij}\). The operator inequality \(P(E_0'-E_1')P\succeq 0\) gives \(a_{0j}-a_{1j}\geq 0\), and the corresponding inequality on the range of \(I-P\) gives \(b_{1j}-b_{0j}\geq 0\).

For any positive semi-definite block matrix $G:=(G_{ab})_{a,b=1}^2$, its square root can be split into two block rows
\begin{equation}
    \sqrt{G} = \begin{pmatrix}
V \\
W
\end{pmatrix}.
\end{equation}
Then, 
\begin{equation}
    G = \sqrt{G}\,(\sqrt{G})^\dagger =
\begin{pmatrix}
VV^\dagger & VW^\dagger \\
WV^\dagger & WW^\dagger
\end{pmatrix}.
\end{equation}
Consequently, 
$G_{11} = V V^\dagger, G_{12} = V W^\dagger$ and $G_{22} = W W^\dagger$.
Setting $A = \sqrt{D_{00}V}, \,\, B= \sqrt{D_{11}}W$, we have
\begin{equation}
    \sqrt{D_{00}} G_{12} \sqrt{D_{11}} = A B^{\dagger}.
\end{equation}
The matrix Cauchy-Schwarz inequality $\|AB^{\dagger} \|_1 \leq \|A\|_2 \|B\|_2$, where $\|A\|_2 = \sqrt{\Tr(AA^\dagger)}$ is the Hilbert-Schmidt norm, gives 
\begin{equation}\label{eq:positive-block-norm}
 \norm{\sqrt{D_{00}}G_{12}\sqrt{D_{11}}}
 \le
 \sqrt{\Tr(D_{00}G_{11})\,\Tr(D_{11}G_{22})}.
\end{equation}
We use the above inequality to bound the positive operators $R_{vv}$ and
$R_{ww}$. Complete positivity applied to the rank-one block
formed from $(w,v)$ bounds the $R_{wv}$ term by
$\sqrt{b_{00}a_{11}}$, namely 
\begin{equation}
    \| \sqrt{D_{00} } R_{wv}\sqrt{D_{11}} \|_1 \leq \sqrt{b_{00}a_{11}}.
\end{equation}
\color{black}
Since $v+w=\chi$, Lemma~\ref{lem:block} gives
\begin{equation}
 \begin{pmatrix}
  R_{ww} & R_{vw}\\
  R_{wv} & R_{vv}
 \end{pmatrix}\succeq0.
\end{equation}
Applying~\eqref{eq:positive-block-norm} to this block matrix yields
\[
 \norm{\sqrt{D_{00}}R_{vw}\sqrt{D_{11}}}
 \le
 \sqrt{\Tr(D_{00}R_{ww})\,\Tr(D_{11}R_{vv})}
 =\sqrt{b_{00}a_{11}}.
\]
\color{black}
Expanding
$R=R_{vv}+R_{ww}+R_{vw}+R_{wv}$ and using the triangle
inequality, yields
\begin{equation}\label{eq:four-split-bound}
 x = \| \sqrt{D_{00}} R \sqrt{D_{11}}\| \le\sqrt{a_{00}a_{11}}+\sqrt{b_{00}b_{11}}
       +2\sqrt{b_{00}a_{11}}.
\end{equation}
Finally,~\eqref{eq:split-weights} implies
$a_{00}\le p_{00}$, $a_{11}\le a_{01}\le p_{01}$,
$b_{00}\le b_{10}\le p_{10}$, and $b_{11}\le p_{11}$.
Finally, substituting this into inequality (\ref{eq:four-split-bound}), we have
\begin{equation}
     x\le\sqrt{p_{00}p_{01}}+\sqrt{p_{10}p_{11}}
       +2\sqrt{p_{01}p_{10}}.
\end{equation}
\end{proof}

\subsection{Tensor products remove the prefactors}

The initial estimate is enough because it can be applied again to
any number of copies. Rescaling balances its first two terms, and
taking tensor powers separates their exponential rates.

\begin{proposition}[Fourth-root bound]\label{prop:four}
Let $\sigma\in\Z(A:B)$. For positive local operators
$X_0,X_1,Y_0,Y_1$, put $D_{ij}=X_i\otimes Y_j$ and
$p_{ij}=\Tr(D_{ij}\sigma)$. Then
\begin{equation}\label{eq:four}
 \norm{\sqrt{D_{00}}\sigma\sqrt{D_{11}}}
 \le(p_{00}p_{01}p_{10}p_{11})^{1/4}.
\end{equation}
The same assertion holds for unnormalized certified outputs,
mixed tensor products, and arbitrary joint local effects across copies.
\end{proposition}

\begin{proof}
First let $R\in\mathcal G^+$ and suppose that all four
weights are positive. In~\eqref{eq:weak-four}, replace
$X_0$ by $tX_0$ and $X_1$ by $t^{-1}X_1$.
The left-hand side is unchanged. Optimizing over $t>0$ gives
\begin{equation}\label{eq:balanced}
 x\le2g+2h,\qquad
 g=(p_{00}p_{01}p_{10}p_{11})^{1/4},\quad
 h=\sqrt{p_{01}p_{10}}.
\end{equation}
Apply this to $R^{\otimes k}$ and the product effects.
Tensor closure and multiplicativity of the trace norm give
$x^k\le2g^k+2h^k$. Taking $k$th roots and letting
$k\to\infty$ yields $x\le\max\{g,h\}$.
Ordinary positivity independently gives
\begin{equation}\label{eq:ordinary-bound}
 x\le\sqrt{p_{00}p_{11}}.
\end{equation}
If $h\le g$, the desired bound already follows.
If $h>g$,~\eqref{eq:ordinary-bound} gives
$x\le g^2/h<g$.

If $p_{00}p_{11}=0$,~\eqref{eq:ordinary-bound} gives $x=0$.
Otherwise, if $p_{01}p_{10}=0$, at most one of the first two
terms in the rescaled form of~\eqref{eq:weak-four} remains.
Letting $t$ tend to zero or infinity makes that term vanish,
so again $x=0$.
Continuity in $R$ for fixed effects extends the result to $\Z$.
The remaining assertions follow from
Proposition~\ref{prop:closure}.
\end{proof}

\section{From shield coherences to secret key}\label{sec:key}

\subsection{Separable operations and shield blocks}

We now apply the four-effect bound to the output of a protocol.
The output systems are $K_AS_A:K_BS_B$, and we write the state
in its key basis as
\begin{equation}\label{eq:shield-expansion}
 \omega=\sum_{a,b,c,e}\ket{ab}\bra{ce}\otimes X_{ab,ce},
 \qquad p_{ab}=\Tr (X_{ab,ab}).
\end{equation}
The operators $X_{ab,ce}$ act on the full shield. In particular,
the coherence relevant to a private-state test is the trace norm
of $X_{ii,jj}$ itself.

\begin{lemma}[Transfer to shield blocks]\label{lem:shield}
Let $R$ be a positive operator, not necessarily normalized,
satisfying~\eqref{eq:four} for all positive local operators.
Let $\Lambda$ be a separable CP map with
$r=\Tr\Lambda(R)>0$, and write $\omega=\Lambda(R)/r$.
For every pair of distinct key labels $i,j$,
\begin{equation}\label{eq:shield}
 \norm{X_{ii,jj}}\le(p_{ii}p_{ij}p_{ji}p_{jj})^{1/4}.
\end{equation}
\end{lemma}

\begin{proof}
Write
$\Lambda(Z)=\sum_\ell(A_\ell\otimes B_\ell)
Z(A_\ell\otimes B_\ell)^\dagger$.
For the key rows
$A_{a,\ell}=(\bra a\otimes I)A_\ell$ and
$B_{b,\ell}=(\bra b\otimes I)B_\ell$, put
$C_{ab,\ell}=A_{a,\ell}\otimes B_{b,\ell}$.
The contribution of Kraus term $\ell$ to the unnormalized
shield block is $C_{ii,\ell}R C_{jj,\ell}^\dagger$.
Rectangular polar decomposition gives
\begin{equation}\label{eq:polar}
 \norm{CRD^\dagger}
 =\norm{\sqrt{C^\dagger C}\,R\sqrt{D^\dagger D}}.
\end{equation}
The outside partial isometries preserve the nonzero singular
values because the middle operator is supported on their
initial spaces.

Apply~\eqref{eq:four} with
$X_0=A_{i,\ell}^\dagger A_{i,\ell}$,
$X_1=A_{j,\ell}^\dagger A_{j,\ell}$ and the analogous
operators on Bob's side. If
$p_{ab}^{(\ell)}=\Tr(C_{ab,\ell}R C_{ab,\ell}^\dagger)$,
then
\begin{align}
 r\norm{X_{ii,jj}}
 &\le\sum_\ell
 \bigl(p_{ii}^{(\ell)}p_{ij}^{(\ell)}
       p_{ji}^{(\ell)}p_{jj}^{(\ell)}\bigr)^{1/4}\notag\\
 &\le
 \left(\sum_\ell p_{ii}^{(\ell)}\right)^{1/4}
 \left(\sum_\ell p_{ij}^{(\ell)}\right)^{1/4}
 \left(\sum_\ell p_{ji}^{(\ell)}\right)^{1/4}
 \left(\sum_\ell p_{jj}^{(\ell)}\right)^{1/4}.
 \label{eq:branch-holder}
\end{align}
The first step uses the triangle inequality, and the second is
four-factor H\"older. Since each population sum equals
$rp_{ab}$, division by $r$ proves the claim.
\end{proof}

The trace norm in~\eqref{eq:shield} also has a direct privacy
interpretation. For a purification
$\ket\Psi=\sum_\alpha\ket\alpha\ket{\eta_\alpha}_{SE}$,
let $\tau_\alpha=\Tr_S\ket{\eta_\alpha}\bra{\eta_\alpha}$
be Eve's subnormalized conditional state. Uhlmann's theorem gives
\begin{equation}\label{eq:corner-fidelity}
 \norm{X_{\alpha,\beta}}
 =\norm{\sqrt{\tau_\alpha}\sqrt{\tau_\beta}}.
\end{equation}
Thus the shield coherence equals the root fidelity of the
corresponding states available to Eve. This is why control of the
untransposed shield block is the relevant quantity.

\subsection{Summing the coherences}

\begin{proposition}[Private-test bound]\label{prop:private}
Let $\omega$ be a normalized state
obeying~\eqref{eq:shield} for every pair $i\ne j$.
Then every twisting satisfies
\begin{equation}\label{eq:private-bound}
 \Tr(\Pi_{d,U}\omega)\le\beta_d,
\end{equation}
where $
\Pi_{d,U}=U(\Phi_d\otimes I_S)U^\dagger.
$
\end{proposition}

\begin{proof}
Let $P=\sum_i p_{ii}$ be the total weight of matching key
labels, and let $Q=\sum_{i\ne j}p_{ij}=1-P$.
Expanding the chosen test and using
$|\Tr(VX)|\le\norm{X}$ for a unitary $V$, gives
\begin{equation}\label{eq:test-expansion}
 \Tr(\Pi_{d,U}\omega)
 \le\frac1d\left(P+\sum_{i\ne j}\norm{X_{ii,jj}}\right).
\end{equation}
The bound holds for the actual twisting; it does not require its
blocks to optimize the individual corner norms simultaneously.
By~\eqref{eq:shield} and Cauchy--Schwarz,
\begin{align}
 \sum_{i\ne j}\norm{X_{ii,jj}}
 &\le
 \sqrt{\left(\sum_{i\ne j}\sqrt{p_{ii}p_{jj}}\right)
       \left(\sum_{i\ne j}\sqrt{p_{ij}p_{ji}}\right)}
 \notag\\
 &\le\sqrt{(d-1)PQ}.
 \label{eq:corner-sum}
\end{align}
For the first factor,
$(\sum_i\sqrt{p_{ii}})^2-P\le(d-1)P$.
For the second, pair the terms $(i,j)$ and $(j,i)$
and use the arithmetic--geometric mean inequality.
It remains to maximize
$P+\sqrt{(d-1)P(1-P)}$ on $[0,1]$.
Its maximum is $(1+\sqrt d)/2$, attained at
$P=(1+d^{-1/2})/2$. Equations~\eqref{eq:test-expansion}
and~\eqref{eq:corner-sum} prove the assertion.
\end{proof}

This estimate expresses the tradeoff behind the result.
A private state needs coherences between matching key labels.
Equation~\eqref{eq:shield} forces those coherences to be
accompanied by weight on mismatching labels.
Optimizing that tradeoff gives a test acceptance strictly below
one, independently of how large the shield is.

\subsection{The operational conclusion}

\begin{proof}[Proof of the private-test and zero-key assertions
in Theorem~\ref{thm:main}]
Proposition~\ref{prop:closure} places every input tensor power in
$\Z$. Proposition~\ref{prop:four},
Lemma~\ref{lem:shield}, and Proposition~\ref{prop:private}
then give~\eqref{eq:main}. Since a private-state test accepts
its target state with probability one,
\begin{equation}\label{eq:distance}
 \frac12\norm{\omega-\gamma_{d,U}}
 \ge1-\Tr(\Pi_{d,U}\omega)\ge1-\beta_d.
\end{equation}
Already for a secret bit,
$\beta_2=(1+\sqrt2)/4<1$. The equivalence of secure-key
distillation and approximate private-state distillation
therefore yields $K_D(\rho)=0$~\cite{general,wilde}.
The membership and NPT assertions for $\rho_*$
are proved in Section~\ref{sec:explicit}.
\end{proof}

We will also use a quantitative version of the passage to classical
key. Let
\begin{equation}\label{eq:ideal-key}
 \kappa_d=\frac1d\sum_{i=0}^{d-1}\ket{ii}\bra{ii}
\end{equation}
be the ideal classical key state. For a completed fixed-length LOCC
protocol, retain the local environments in the shield before the final
key readout. If $\vartheta_{K_AK_BE}$ is its classical-key
output and $\omega$ its retained quantum output, then for each
state $\theta_E$ there is a private-state test with
\begin{equation}\label{eq:ccq-fidelity}
 f(\vartheta,\kappa_d\otimes\theta_E)^2
 \le\Tr(\Pi_{d,U}\omega),
 \qquad f(\rho,\sigma)=\norm{\sqrt\rho\sqrt\sigma}.
\end{equation}
Appendix~\ref{app:classical} proves this statement, keeping
the public transcript with Eve and all private environments local.
Consequently, the usual trace-distance error obeys
\begin{equation}\label{eq:ccq-error}
 \frac12\norm{\vartheta-\kappa_d\otimes\theta_E}
 \ge1-\sqrt{\Tr(\Pi_{d,U}\omega)}.
\end{equation}
For an input in $\Z$, the right-hand side is at least
$1-\sqrt{\beta_d}$.
Since
\begin{equation}\label{eq:bits}
 \beta_{2^m}=\tfrac12(2^{-m}+2^{-m/2}),
\end{equation}
the acceptance tends to zero exponentially with the target key
length. This statement refers to the unconditional output of a
completed protocol. Theorem~\ref{thm:main} separately controls
normalized nonzero branches.

\section{An explicit NPT state}\label{sec:explicit}

We now use the state and rational certificate of
Ref.~\cite{faisal}. On $\mathbb C^8$, define
\begin{equation}\label{eq:explicit-maps}
 L(X)=5\Tr(X)I-3X^T,\qquad
 M(X)=18\Tr(X)I+37X,
 \qquad \ket u=\sum_{i=0}^7\ket{ii}.
\end{equation}
Their Choi matrices are
$J_L=5I-3F\succeq0$ and
$J_M=18I+37\ket u\bra u\succeq0$, so both maps are CP.
They are not coCP: $J_{L\circ T}=5I-3\ket u\bra u$
has eigenvalue $-19$ on $u$, and
$J_{M\circ T}=18I+37F$ has eigenvalue $-19$ on the
antisymmetric subspace.

The output on the ancestor is
\begin{equation}\label{eq:explicit-output}
 (L\otimes M)(\ket u\bra u)
 =851I-111F=37(23I-3F).
\end{equation}
Since $\Tr I=64$ and $\Tr F=8$, normalization gives
$\rho_*$ in~\eqref{eq:rho-star}. Its eigenvalues are
$20/1448$ on the symmetric subspace and $26/1448$ on
the antisymmetric subspace. It is therefore a state.
On the other hand,
\begin{equation}\label{eq:npt}
 \rho_*^{\Gamma_B}
 =\frac{23I-3\ket u\bra u}{1448},\qquad
 \rho_*^{\Gamma_B}\ket u=-\frac1{1448}\ket u.
\end{equation}
Thus $\rho_*$ is NPT.

To prove membership in the certified family, order the four Choi
factors as $A_{\mathrm{in}},B_{\mathrm{in}},
A_{\mathrm{out}},B_{\mathrm{out}}$.
The Choi matrix of $L\otimes M$ is
\begin{equation}\label{eq:explicit-J}
 J=(5I-3F_{13})(18I+37P_{24}),\qquad
 P_{24}=\ket u_{24}\bra u.
\end{equation}
Appendix~\ref{app:verify} specifies a rational Hermitian
matrix $Q$ by 24 integer coefficients and verifies the
following conditions.

\begin{proposition}[Exact certificate]\label{prop:explicit}
For $J$ in~\eqref{eq:explicit-J} and $Q$
in~\eqref{eq:Q},
\begin{equation}\label{eq:explicit-cert}
 J+Q\succeq0,\qquad J-Q\succeq0,\qquad
 Q^{\Gamma}\succeq0,\qquad
 (J-Q)(\ket u\otimes I)=0.
\end{equation}
The ranks of these three positive matrices are respectively
$4096$, $4032$, and $4096$.
Moreover,
\begin{equation}\label{eq:exact-kernel}
 \ker(J-Q)=\ran(\ket u\otimes I).
\end{equation}
\end{proposition}

The finite verification is described in
Appendix~\ref{app:verify} and reproduced by the accompanying
Python program. It uses integer congruence elimination rather than
numerical eigenvalue tolerances. The analytic role of the result is
immediate: Lemma~\ref{lem:matching} turns~\eqref{eq:explicit-cert}
into a certified representation of
\eqref{eq:explicit-output}. Hence $\rho_*\in\mathcal G
\subset\Z$, completing Theorem~\ref{thm:main}.
Since the local CP/coCP subfamily is PPT, $\rho_*$
also proves that $\Z$ strictly extends its trace-norm closure.

\section{Relative entropy and a strong converse}\label{sec:entropy}

\subsection{A relative-entropy bound on key}

The family $\Z$ can serve as a reference set for states that
need not themselves be certified. Define
\begin{align}
 \Ecert(\tau)
 &=\inf_{\sigma\in\Z(A:B)}D(\tau\Vert\sigma),\label{eq:Ecert}\\
 a_n(\tau)
 &=\inf_{\sigma_n\in\Z(A^n:B^n)}
       D(\tau^{\otimes n}\Vert\sigma_n),\label{eq:an}
\end{align}
where
$D(\rho\Vert\sigma)=\Tr\rho(\log\rho-\log\sigma)$
when $\supp\rho\subseteq\supp\sigma$, and is infinite
otherwise. Because $\Z$ contains the maximally mixed
state, $a_n$ is finite. Mixed tensor closure and additivity of
relative entropy give $a_{n+k}\le a_n+a_k$, so
\begin{equation}\label{eq:regularization}
 \Ecert^\infty(\tau)
 =\lim_{n\to\infty}\frac{a_n(\tau)}n
 =\inf_{n\ge1}\frac{a_n(\tau)}n.
\end{equation}

\begin{theorem}[Relative-entropy key bound]\label{thm:entropy}
For every finite-dimensional bipartite state $\tau$,
\begin{equation}\label{eq:key-bound}
 K_D(\tau)\le2\Ecert^\infty(\tau)\le2\Ecert(\tau).
\end{equation}
\end{theorem}

\begin{proof}
Consider a completed trace-preserving $n$-copy LOCC protocol
with a target of $m_n=\log d_n$ key bits. Apply a private-state
test to its retained quantum output. Let $p_n$ be the
acceptance on $\tau^{\otimes n}$, and let $q_n$ be the
acceptance when the same protocol acts on
$\sigma_n\in\Z(A^n:B^n)$.
Closure under separable operations and
Theorem~\ref{thm:main} give
\begin{equation}\label{eq:q-bound}
 q_n\le\beta_{d_n}\le d_n^{-1/2}=2^{-m_n/2}.
\end{equation}
Data processing to this binary test yields
\begin{align}
 D(\tau^{\otimes n}\Vert\sigma_n)
 &\ge
 p_n\log\frac{p_n}{q_n}
 +(1-p_n)\log\frac{1-p_n}{1-q_n}\notag\\
 &\ge \frac{p_n m_n}{2}-h_2(p_n),
 \label{eq:binary-relative-entropy}
\end{align}
where $h_2$ is binary entropy.
For a successful key-distillation sequence, the corresponding
tests may be chosen with $p_n\to1$, by
\eqref{eq:ccq-fidelity}. Infimize over $\sigma_n$, divide by
$n$, and use~\eqref{eq:regularization}. This proves the first
inequality in~\eqref{eq:key-bound}. Tensor powers of a
one-copy reference prove the second.
\end{proof}

The coefficient two comes from the decay
$q_n\lesssim d_n^{-1/2}$ in~\eqref{eq:q-bound}.
Replacing the local CP/coCP reference family by $\Z$
can only decrease the relative-entropy infimum, because the
reference set becomes larger. A strict numerical improvement for
an individual input would require a separate comparison.

\subsection{Exponential failure above the regularized bound}

Relative-entropy data processing alone gives the rate bound.
To prove that the error tends to its maximal value above that
rate, we use a R\'enyi divergence and fix the reference block
before choosing its order.

\begin{theorem}[Strong converse]\label{thm:strong}
Let $\tau$ be finite dimensional. Consider completed,
trace-preserving, fixed-length LOCC protocols acting on
$\tau^{\otimes n}$, with finite local output systems at each
$n$ and target key dimensions $d_n$. If
\begin{equation}\label{eq:strong-rate}
 \liminf_{n\to\infty}\frac{\log d_n}{n}
 >2\Ecert^\infty(\tau),
\end{equation}
there are constants $c>0$ and $C_0<\infty$ such that,
for all sufficiently large $n$, every private-state test on
the retained output $\omega_n$ satisfies
\begin{equation}\label{eq:exponential-test}
 \Tr(\Pi_{d_n,U}\omega_n)\le2^{-cn+C_0}.
\end{equation}
For the classical-key output $\vartheta_n$ and its Eve
marginal $\vartheta_{E,n}$, the secrecy-and-correctness
error therefore obeys
\begin{equation}\label{eq:strong-error}
 \frac12\norm{\vartheta_n-\kappa_{d_n}\otimes\vartheta_{E,n}}
 \ge1-2^{-cn/2+C_0/2}.
\end{equation}
\end{theorem}

\begin{proof}
For $1<\alpha\le2$, the Petz R\'enyi divergence
\begin{equation}\label{eq:petz}
 D_\alpha(\rho\Vert\sigma)
 =\frac1{\alpha-1}\log\Tr(\rho^\alpha\sigma^{1-\alpha})
\end{equation}
is additive and obeys data processing~\cite{renyi}.
Take a full-rank reference $\sigma_n\in\Z$ and
apply the protocol and a binary private-state test.
The classical R\'enyi expression is at least
$p_n^\alpha q_n^{1-\alpha}$ before taking the logarithm.
Using~\eqref{eq:q-bound} gives
\begin{equation}\label{eq:renyi-test}
 p_n\le
 2^{-\frac{\alpha-1}{\alpha}
 [m_n/2-D_\alpha(\tau^{\otimes n}\Vert\sigma_n)]}.
\end{equation}
The case $p_n=0$ is immediate.

Choose $R_0$ strictly between $2\Ecert^\infty(\tau)$
and the liminf in~\eqref{eq:strong-rate}.
By~\eqref{eq:regularization}, there are a finite $k$
and $\sigma_k\in\Z(A^k:B^k)$ with
\begin{equation}\label{eq:finite-block-gap}
 \frac1kD(\tau^{\otimes k}\Vert\sigma_k)<\frac{R_0}{2}.
\end{equation}
Mix $\sigma_k$ with a sufficiently small positive amount
of the maximally mixed state. Convexity preserves membership
in $\Z$, the reference becomes full rank, and the strict
inequality remains valid. For this fixed block and reference,
$D_\alpha\to D$ as $\alpha\downarrow1$.
We can therefore choose a fixed $1<\alpha\le2$ such that
\begin{equation}\label{eq:renyi-gap}
 \frac{A_\alpha}{k}<\frac{R_0}{2},
 \qquad
 A_\alpha=D_\alpha(\tau^{\otimes k}\Vert\sigma_k).
\end{equation}

Write $n=\ell k+r$, with $0\le r<k$, and let $\mu$
be the one-copy maximally mixed state. Use
\begin{equation}\label{eq:block-reference}
 \widehat\sigma_n=\sigma_k^{\otimes\ell}\otimes\mu^{\otimes r}
 \in\Z(A^n:B^n).
\end{equation}
By additivity,
\begin{equation}\label{eq:block-renyi}
 D_\alpha(\tau^{\otimes n}\Vert\widehat\sigma_n)
 =\ell A_\alpha+rD_\alpha(\tau\Vert\mu)
 \le \frac{nA_\alpha}{k}+C
\end{equation}
for a constant $C$ independent of $n$.
Eventually $m_n\ge nR_0$. Substitute
\eqref{eq:block-renyi} into~\eqref{eq:renyi-test} and choose
\begin{equation}\label{eq:exponent}
 c=\frac{\alpha-1}{\alpha}
       \left(\frac{R_0}{2}-\frac{A_\alpha}{k}\right)>0.
\end{equation}
This proves~\eqref{eq:exponential-test}, uniformly in the
private-state test. Finally, apply~\eqref{eq:ccq-fidelity}
to the test supplied for
$\theta_E=\vartheta_{E,n}$ and then
\eqref{eq:ccq-error}. This gives~\eqref{eq:strong-error}.
\end{proof}

The order of choices in this proof is essential. The block length,
the reference, and its full-rank perturbation are fixed before
$\alpha$ approaches one. The argument therefore uses continuity
in a fixed finite-dimensional problem, without interchanging a
R\'enyi limit with regularization. The rate statement concerns
unconditional completed protocols; it assigns no unconditional
rate to an arbitrarily rare conditioned branch.

\section{Discussion and conclusions}\label{sec:discussion}

We have proved that the NPT Werner state $\rho_*$ has zero
distillable secret key. Its negative partial transpose certifies
entanglement, yet collective local processing and public
communication cannot convert it into a private key. For this
state, the result extends the obstruction to Bell-pair
distillation established in Ref.~\cite{faisal}. The extension
addresses a distinct requirement: bound entanglement can contain
secret key, so ruling out Bell pairs alone does not rule out
the privacy protected by a shield~\cite{secure,general}.

The proof controls the operators on that shield, independently
of its finite dimension. Coherences between matching key
labels require weight on mismatching labels, and this tradeoff
bounds the acceptance of every private-state test by $\beta_d$.
The auxiliary certificate makes the tradeoff stable under tensor
products and separable processing. This gives a closed convex
family containing both the earlier local CP/coCP
family~\cite{openai} and the explicit NPT example.
Any finite collection of its members still has zero key when
processed jointly. This stability is a property of the certified
family; it is not assumed for arbitrary zero-key states.

For general inputs, the same family supplies reference states
for a relative-entropy key bound. Rates above twice the
regularized relative entropy have exponentially vanishing
private-test acceptance. For completed fixed-length protocols,
the secrecy-and-correctness error consequently tends to one.
The enlarged reference family can lower the relative-entropy
infimum. Establishing a strict improvement over other key bounds
for a particular input requires a separate quantitative comparison.

The explicit example uses the finite rational certificate
supplied by Faisal~\cite{faisal}. Appendix~\ref{app:verify}
gives its verification in exact integer arithmetic, with a
reproducible program supplied among the ancillary files.
Once these finite matrix conditions are checked, the conclusions
for arbitrary copy number, finite shield dimension, and
separable processing follow analytically.

\section*{Acknowledgements}

KH thanks Pawe{\l} Mazurek for bringing Faisal’s recent result to his attention.

KH and MW acknowledge support from the National Science Centre, Poland, under OPUS~25 Grant No.~2023/49/B/ST2/02468.

VB acknowledges support from VILLUM FONDEN via the QMATH Centre of Excellence (Grant No.~10059), and the Novo Nordisk Foundation (Grant No.~NNF20OC0059939, “Quantum for Life”). TT acknowledges support from a Postdoc Scholarship on Quantum Algorithms or Quantum Software from the Danish e-infrastructure Consortium (DeiC).

The authors acknowledge the use of ChatGPT (OpenAI, GPT-6 Astra)
and Gemini (Google) for assistance with checking the mathematical
arguments and with language and structural editing of the first
version of the manuscript. The authors take full responsibility
for the scientific content and the final text.

\section*{ Code availability}

The code used in this study is available from the authors upon request.

\clearpage
\appendix
\section*{Appendices}

\section{The rational certificate and its exact verification}
\label{app:verify}

\subsection{Coefficients and tensor order}

Use the tensor order
$A_{\mathrm{in}},B_{\mathrm{in}},A_{\mathrm{out}},B_{\mathrm{out}}$,
with each factor of dimension eight.
For a permutation $\pi\in S_4$, define
\begin{equation}\label{eq:permutation}
 (U_\pi)_{r,c}
 =\prod_{i=1}^4\delta_{r_i,c_{\pi(i)}}.
\end{equation}
The certificate of Ref.~\cite{faisal} is
\begin{equation}\label{eq:Q}
 Q=\frac1{40}\sum_{\pi\in S_4}m_\pi U_\pi^{\Gamma_2}.
\end{equation}
Here $\Gamma_2$ transposes the second factor only.
The input transpose $\Gamma=\Gamma_{12}$
instead transposes both input factors. Table~\ref{tab:coeff}
lists every coefficient, with permutations written as
$\pi(1)\pi(2)\pi(3)\pi(4)$.

\begin{table}[htbp]
\centering
\begin{tabular}{rr@{\qquad}rr@{\qquad}rr}
\toprule
$\pi$ & $m_\pi$ & $\pi$ & $m_\pi$ & $\pi$ & $m_\pi$\\
\midrule
1234 & 88  & 2314 & -288 & 3412 & 612\\
1243 & 77  & 2341 & 128  & 3421 & -756\\
1324 & 16  & 2413 & -525 & 4123 & 128\\
1342 & 24  & 2431 & 760  & 4132 & 760\\
1423 & 24  & 3124 & -288 & 4213 & -876\\
1432 & 16  & 3142 & -525 & 4231 & 2060\\
2134 & 380 & 3214 & 884  & 4312 & -756\\
2143 & 40  & 3241 & -876 & 4321 & -172\\
\bottomrule
\end{tabular}
\caption{Integer coefficients in~\eqref{eq:Q}. Reading down
each pair of columns, from left to right, gives lexicographic
permutation order.}
\label{tab:coeff}
\end{table}

\subsection{An exact finite check}

The accompanying program
\texttt{verify\_npt\_certificate.py} constructs
$40J$ and $40Q$ directly from
\eqref{eq:explicit-J}, \eqref{eq:permutation}, and
\eqref{eq:Q}. It checks Hermiticity entrywise and computes
the input partial transpose explicitly. The three matrices
$40(J+Q)$, $40(J-Q)$, and
$40Q^{\Gamma}$ have integer entries.

For each matrix, the program finds the connected components
of its nonzero-entry graph and permutes it into the
corresponding diagonal blocks. Each test has 708 components,
with size distribution
\begin{equation}\label{eq:block-histogram}
 1^{[56]},\qquad 3^{[336]},\qquad 6^{[280]},\qquad
 22^{[8]},\qquad 42^{[28]},
\end{equation}
where the bracketed exponent denotes the number of components.
Their dimensions sum to 4096. Identical ordered component
matrices are checked once with their multiplicities retained;
this leaves 40 distinct matrices for each test.

Positivity is decided by exact integer congruence elimination.
For a symmetric integer block
\begin{equation}\label{eq:schur-block}
 B=\begin{pmatrix}p&v^T\\v&B_{22}\end{pmatrix},
\end{equation}
a negative leading entry rejects positivity. If $p=0$,
positivity requires the entire leading row and column to be
zero; they are then removed. If $p>0$, replace the trailing
block by
\begin{equation}\label{eq:integer-schur}
 B_{\mathrm{next}}=pB_{22}-vv^T.
\end{equation}
This is a positive multiple of the Schur complement, so
$B\succeq0$ if and only if $B_{\mathrm{next}}\succeq0$,
and the rank increases by one. Dividing
$B_{\mathrm{next}}$ by the positive greatest common
divisor of its entries controls integer growth without changing
positivity or rank. Every multiplication in this elimination
uses Python arbitrary-precision integers.

The exact results are shown in Table~\ref{tab:verify}.
Each matrix has 66,424 nonzero entries. No floating-point
eigenvalues or tolerance thresholds enter the decisions.

\begin{table}[htbp]
\centering
\begin{tabular}{lrrrr}
\toprule
Matrix & Dimension & Rank & Nullity & Largest entry (bits)\\
\midrule
$40(J+Q)$ & 4096 & 4096 & 0  & 513\\
$40(J-Q)$ & 4096 & 4032 & 64 & 457\\
$40Q^{\Gamma}$ & 4096 & 4096 & 0 & 151\\
\bottomrule
\end{tabular}
\caption{Results of integer elimination. All three matrices
are positive semidefinite. The last column records the largest
matrix entry encountered, including Schur-complement entries
before gcd reduction.}
\label{tab:verify}
\end{table}

For matching, form the sparse embedding
$V=\ket u\otimes I$ with entries
$V_{(t,t,k,l),(k,l)}=1$.
The program verifies
\begin{equation}\label{eq:matching-residual}
 40(J-Q)V=0,\qquad V^\dagger V=8I_{64}.
\end{equation}
Thus $\ran V$ is a 64-dimensional subspace of
$\ker(J-Q)$. The computed nullity is also 64, proving
\eqref{eq:exact-kernel}.

Two independent constructions check the tensor conventions.
First, $J$ is also built from $J_L\otimes J_M$ by
explicitly regrouping its factors. Second,
$40Q^{\Gamma_{12}}$ is also built as
$\sum_\pi m_\pi U_\pi^{\Gamma_1}$.
Both constructions agree entrywise with the primary matrices.
The program additionally sums the actions of $L$ and $M$
on matrix units to verify~\eqref{eq:explicit-output}, and
checks the negative eigenpair in~\eqref{eq:npt}.
Together these computations prove
Proposition~\ref{prop:explicit}.

\subsection{Reproducibility}

The ancillary files contain the verifier and its output record,
\texttt{verification\_results.json}.
With Python 3.10 or later, NumPy, and SciPy installed, run
the following command from their directory:
\begin{verbatim}
python3 verify_npt_certificate.py
\end{verbatim}
The program writes a fresh JSON record beside itself; the
\texttt{--output} option selects a different destination.
The record contains the coefficient list, software versions,
the script hash, exact ranks, component multiplicities,
block hashes, and elimination pivot signs.
NumPy and SciPy are used for integer sparse assembly and graph
components; all positivity and rank decisions use arbitrary-precision
integer arithmetic. The companion files are supplied separately
from the PDF.

\section{From classical key to a private-state test}
\label{app:classical}

Let $\vartheta_{K_AK_BE}$ be the classical output of a completed
fixed-length protocol with $d$ key values. Retain all private local
environments in a shield $S=A'B'$, allowing additional local pure
ancillas, and denote the quantum output before the final key readout by
$\omega_{K_AK_BS}$. Eve's system $E$ includes the input purification and
the entire public transcript. We prove that for every state $\theta_E$
there exists a twisting $U$ such that
\begin{equation}
 f(\vartheta,\kappa_d\otimes\theta_E)^2
 \leq \operatorname{Tr}(\Pi_{d,U}\omega),
 \qquad
 \Pi_{d,U}=U(\Phi_d\otimes I_S)U^\dagger,
 \label{eq:classical-key-private-test}
\end{equation}
where $\kappa_d=d^{-1}\sum_i|ii\rangle\langle ii|$,
$\Phi_d=|\Phi_d\rangle\langle\Phi_d|$,
$|\Phi_d\rangle=d^{-1/2}\sum_i|ii\rangle$, and
$f(X,Y)=\|\sqrt X\sqrt Y\|_1$.

\begin{proof}
Use the coherent realization of the protocol from
\cite[Sec.~IX-A and Theorem~5]{secure}: purify local operations and
retain their environments locally, while copying every public message
to Eve. The resulting pure state can be written as
\[
 |\Psi\rangle_{K_AK_BSE}
 =\sum_{i,j}|ij\rangle|\psi_{ij}\rangle_{SE},
 \qquad
 \omega=\operatorname{Tr}_E|\Psi\rangle\langle\Psi|.
\]
Its Alice--Bob marginal is obtainable by LOCC, since tracing Eve's
message registers implements ordinary classical communication.
The vectors $|\psi_{ij}\rangle$ are unnormalized, and the final key
measurement gives
$\vartheta=\sum_{i,j}|ij\rangle\langle ij|\otimes
\operatorname{Tr}_S|\psi_{ij}\rangle\langle\psi_{ij}|$.

Fix $\theta_E$ and a purification $|\phi\rangle_{SE}$, enlarging $S$
by a local pure ancilla if necessary. Set
$\tau_i=\operatorname{Tr}_S|\psi_{ii}\rangle\langle\psi_{ii}|$.
Uhlmann's theorem gives unitaries $U_i$ on $S$, with phases chosen so that
\[
 \langle\psi_{ii}|(U_i\otimes I_E)|\phi\rangle
 =f(\tau_i,\theta_E)\geq0.
\]
Define
\[
 |\Gamma\rangle
 =\frac{1}{\sqrt d}\sum_i|ii\rangle
       (U_i\otimes I_E)|\phi\rangle.
\]
For the twisting with diagonal blocks $U_{ii}=U_i$
(and $U_{ij}=I_S$ for $i\ne j$), its marginal is the private state
$\gamma=U(\Phi_d\otimes\sigma_S)U^\dagger$,
where $\sigma_S=\operatorname{Tr}_E|\phi\rangle\langle\phi|$.
The direct-sum formula for fidelity yields
\[
 \langle\Psi|\Gamma\rangle
 =\frac{1}{\sqrt d}\sum_i f(\tau_i,\theta_E)
 =f(\vartheta,\kappa_d\otimes\theta_E).
\]
Since $(\Pi_{d,U}\otimes I_E)|\Gamma\rangle=|\Gamma\rangle$,
Cauchy--Schwarz gives
\[
 f(\vartheta,\kappa_d\otimes\theta_E)^2
 =\bigl|\langle\Psi|(\Pi_{d,U}\otimes I_E)|\Gamma\rangle\bigr|^2
 \leq\langle\Psi|(\Pi_{d,U}\otimes I_E)|\Psi\rangle
 =\operatorname{Tr}(\Pi_{d,U}\omega).
\]
\end{proof}

Consequently, the fidelity--trace-distance inequality gives
\begin{equation}
 \frac12\|\vartheta-\kappa_d\otimes\theta_E\|_1
 \geq 1-f(\vartheta,\kappa_d\otimes\theta_E)
 \geq 1-\sqrt{\operatorname{Tr}(\Pi_{d,U}\omega)}.
 \label{eq:classical-key-trace-error}
\end{equation}
Here $U$ may depend on $\theta_E$. A bound
$\operatorname{Tr}(\Pi_{d,U}\omega)\leq\beta_d$ valid for every twisting
therefore implies
$\inf_{\theta_E}\frac12\|\vartheta-\kappa_d\otimes\theta_E\|_1
\geq 1-\sqrt{\beta_d}$.

\end{document}